\documentclass{aa}

\usepackage{graphicx}

\begin{document}

\title{Detection of compact objects by means of gravitational lensing
in binary systems}

\author{G.M.~Beskin\inst{1,3} \and A.V.~Tuntsov\inst{1,2}}
\authorrunning{G.M.~Beskin \and A.V.~Tuntsov}
\titlerunning{Gravitational lensing in compact binaries}
\offprints{G. Beskin}

\institute{
Special Astrophysical Observatory, Nizhnij Arkhyz,
	      Karachaevo-Cherkesia, 369167, Russia \\
\email{beskin@sao.ru}
\and  Sternberg Astronomical Institute of the Moscow State University,
Moscow, Russia \\
\email{tyomich@sai.msu.ru}
\and Isaac Newton Institute of Chile, SAO Branch programme}

\date{Received 27 November 2001; Accepted 19 June 2002 }

\abstract{
We consider the gravitational magnification of light for binary
systems containing two compact objects: white dwarfs, a white dwarf and
a neutron star or a white dwarf and a black hole. Light curves of the flares
of the white dwarf caused by this effect were built in analytical
approximations and by means of numerical calculations. We estimate the
probability of the detection of these events in our Galaxy for different
types of binaries and show that gravitational lensing provides a tool for
detecting such systems. We propose to use the facilities of the Sloan
Digital Sky Survey (SDSS) to search for these flares. It is possible to
detect several dozen compact object pairs in such a programme over 5
years. This programme is apparently the best way to detect stellar mass
black holes with open event horizons.
}

\maketitle

\section{Introduction}

One of the most important manifestations of gravitational lensing is the
visible variability of the astrophysical objects whose emission is
affected by the gravitation field of the lens. This effect becomes observable
over a reasonable time period, when the velocities of relative motions of the
observer, the lens and the source are great enough, i.e. their mutual
location changes rapidly. Many cases of this kind of variability have
been examined: from light variation of distant quasars to outburst of
stars caused by the influence of planets (see, for instance,
Zakharov 1997). The brightness variation of the binary system components
caused by gravitational lensing was been first
considered by Ingel (1972, 1974) and Maeder (1973). They noted that
repetition and small characteristic times of the effect make it one
of the most accessible to detection and detailed study. It has also been
shown (Maeder 1973) that the large amplitudes of brightness variations
($0\fm 1-0\fm 5$) can be expected in binary systems consisting of
compact objects -- white dwarfs, neutron stars, black holes. In 
fact, only in these cases does the Einstein-Hvol'son radius turn out to be
smaller than the lens star radius, and the image of the radiation source is
 practically not overlapped by it. The latter circumstance makes the
gravitational lensing influence in binary systems an effective means
for the detection and detailed investigation of compact objects with the aid
of photometric methods alone. Their permeability is by $2^{\rm m}-3^{\rm m}$
higher (as compared to spectroscopy), and, therefore, the number of stars
amenable to study is an order of magnitude larger. Here the light curve
analysis determines all the parameters of the binary system in
full analogy to this task for eclipsing binary systems (with
allowance made for its singularity) (see, for example, Goncharsky et al.
1985).

A complete set of data for compact object binaries and their statistical
analysis gives a possibility to test binary evolution theories and to
investigate the last stages of single star evolution as well. We note that
only 10 double white dwarfs were detected during the last 10 years 
(Maxted~et~al. 2000) and about 40 pairs of a white dwarf and a neutron star --
during 25 years (Thorsett \& Chakrabarty 1999). We further note that by
means of gravitational lensing it is possible to detect compact object
binaries at the same rate (see Section 5). Moreover, this may help to
discover binaries containing neutron stars with a low magnetic field
(not pulsars). Study of these systems in combination with binary star
evolution theories gives a possibility to test the detailed cooling models
for white dwarfs and neutron stars and the equation of state of the latter
(Nelemans~et~al. 2001; Yakovlev~et~al. 1999).

Studying compact object pairs is very important to solve a number of
astrophysical problems. We present below several such examples.

The "Standard candle" of modern cosmology, type Ia supernovae, apparently
arises from merging of double CO white dwarfs (Webbink 1984; Iben \& Tutukov
1984), which have not been detected at this time.

Close double compact objects have to contribute a significant part of the
gravitational wave signal at low frequencies. Thus, white dwarf pairs may be
a source of the unresolved noise (Evans~et~al. 1987; Grischuk~et~al. 2001).
Statistical properties of the closest double compact objects in principle 
determine parameters of the gravitational wave signal. 

Binary systems consisting of a white dwarf and a neutron star (a pulsar)
are unique laboratories for high-precision tests of general relativity.
To date, post-Keplerian general relativistic parameters have been measured
in four such pairs (Thorsett \& Chakrabarty 1999). This problem may be solved
very  easily for binaries with gravitational self-lensing since they are
observed nearly edge-on.

It seems that the detection of black holes forming pairs with white dwarfs 
might be an extraordinarily important result of the search for gravitational
lensing in compact object binaries. Despite the opinion of most researchers,
in a certain sense black holes have not been discovered so far.
Only observational data on the behaviour of matter close to the event horizon
showing its presence may testify that a black hole is identified
(Damour 2000). There is indirect evidence of the presence of black
holes in X-ray binaries and galactic cores based on the "mass-size" relation
expected for them (Cherepashchuk 2001). The horizon neighbourhood
is seen neither in X-ray binaries nor in the cores of active galaxies
because they are screened out by the accreted gas (the accretion rates are 
very high). This means that only black holes accreting usual interstellar plasma
at low rates of $10^{-14}-10^{-15} M_{\odot}$/year can be recognized as
objects without a surface and with an event horizon (Shvartsman 1971).
The black hole companions of white dwarfs in the binaries detected by means 
of gravitational lensing could be the best objects for horizon study and 
tests of general relativity in the strong field limit (Damour 2000).
It is very easy to measure the mass and size of such a black hole due 
to binary edge-on orientation and investigate the radiation of gas near the
horizon. This matter is accreted from interstellar medium only because 
its transfer from the white dwarf is absent.

The gravitational lensing effects in binary systems consisting of compact
components were studied previously in several papers (Maeder 1973;
Gould 1995; Qin~et~al. 1997). However, the authors did not go further
than estimation of the probability of detecting the effect of orientation of
the binary system (often underestimating it by a factor of 2-3), which turned
out to be very low. Their general conclusion is that the effect cannot be
actually recorded. Nevertheless, the data accumulated by now
on the evolution of binary systems and their parameters make it possible to
define with higher accuracy the probability of detection of the brightness
enhancement in binary systems, to find the expected number of such flares
and to propose in the final analysis the strategy for their search based
on present-day facilities. Our paper is devoted to performing these tasks.

We examine in Section 2 the distinguishing features of light magnification in
the systems consisting of two white dwarfs, a white dwarf and a neutron
star, and also a white dwarf and a black hole. In Section 3 we derive
probabilities of recording the effect for its different amplitudes in
all three cases. In Section 4 the numbers of systems of different types
which may be detected by means of gravitational lensing are estimated. In
Section 5 we discuss the possibilities of the quest for brightness magnification
in such systems with the aid of the telescope and equipment used in
the survey SDSS (York et al. 2000).

\section{Light curves in gravitational lensing}

We will consider three
types of binary systems consisting of\\
a) two white dwarfs with masses of $0.7 M_{\odot}$ (sometimes $0.5
M_{\odot})$,\\
b) a white dwarf ($M_{WD} = 0.7 M_{\odot}$) and a neutron star
($M_{NS} = 1.4 M_{\odot}$), \\
c) a white dwarf ($M_{WD} = 0.7 M_{\odot}$) and a black hole ($M_{BH} =
10 M_{\odot}$).\\
It is clear that both components of a pair play the part of a gravitational
lens alternately. However, we will mention at once that the source
of radiation will always be the white dwarf to which the approximation
of a uniformly luminous disk can be applied. The amplitude, the shape and
the duration of flares of the lensed object are defined by the relationships
between the parameters of the binary system, the major semiaxis $a$, the masses
and radii of the components and the orbital plane orientation.

The luminosity magnification of a uniformly radiating disk in gravitational
lensing was analysed in detail in the papers by Refsdal (1964), Liebes (1964),
Byalko (1969), Ingel (1972, 1974), Maeder (1973) and Agol (2002).
On account of the conservation of the surface brightness, the task is
reduced in the final analysis to a study of variations of the source image
area with allowance made for its occultation by an opaque lens, as the source
and lens are moving with respect to the observer (in our case -- the orbital
motion around the mass centre of the binary system).

\begin{figure}[b]
\centering\includegraphics[width=6.5cm,clip=, angle=270 %
]{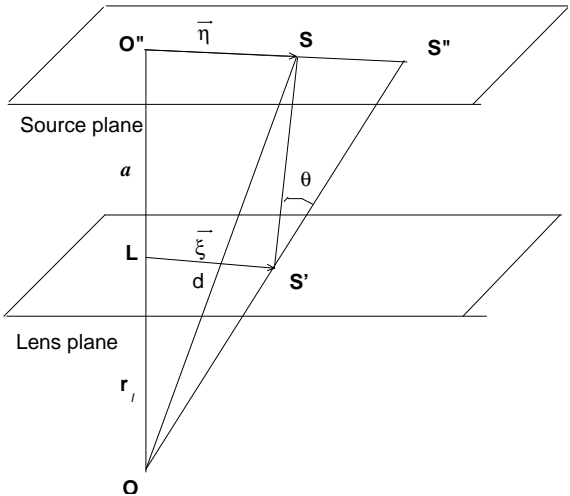}
\caption{Schematic view of gravitational lensing.}
\end{figure}

Let $r_{l}$ be the distance from the observer to the gravitational lens,
$a$ is the binary system orbit semiaxis (for the sake of simplicity we
consider only circular motions), $M_l$ is the lens mass, $r_g = 2GM_{l}/c^2$
is its gravitational radius, $R_l$ is the lens radius, $R_s$ is the source
radius, $M_s$ is the source mass.

In our case for the light path we can use with high accuracy an
approximation of a broken line, the angle between two sections of which is
inversely proportional to the impact parameter.

The gravitational lens equation for a point-like source (see, for instance,
Zakharov 1997) is
\begin{equation}
\overline\eta = \frac{r_l + a}{r_l}\overline\xi -
a\overline\theta(\overline\xi),
\end{equation}
where $\overline\eta $ and $\overline\xi$ are the vectors
that characterize the location of the points of intersection of the
light ray with the planes of the source and the lens, respectively,
$\overline\theta (\overline\xi )$ is the angle of light
deflection (see Fig.~1). In our designations $\overline\theta
(\overline\xi ) = \frac{2 r_g}{\xi}\cdot\frac{\overline\xi}
{\xi}$. The main parameter of the task is the Einstein-Hvol'son radius
$r_{e} = \sqrt{2 r_g\frac{a r_l}{r_l + a}}$, the radius of the ring -
the image of the point source, lying on the "observer -- lens" axis in the
lens plane. In our case $r_l\gg a$ and
\begin{equation}
r_{e} = \sqrt{2 r_g a} 
\end{equation}
or, in solar units,
$$
 r_e=3\cdot 10^{-3}M^{1/2}a^{1/2}.
$$
Equation (1) has two solutions and hence the source images are two points
in the lens plane with coordinates:
\begin{equation}
\xi_{1,2} = \frac{d}{2}\left (1\pm \sqrt{1 +\frac{4 r_{e}^{2}}{d^2}}\right ),
\end{equation}
where $d=|\overline\eta |$ is the "lens--source" vector projection
onto the lens plane, i.e. displacement of the source off the
"observer--lens" axis.

For the magnification coefficient $K$, i.e. the ratio of the sum of
brightness of the two images to the point-like source brightness,
we have (Zakharov 1997)
\begin{equation}
K =\left |\frac{\xi d\xi}{\eta d\eta}\right |_1+ \left |\frac{\xi d\xi}
{\eta d\eta}\right |_2
  = \frac{1}{2}\left (\sqrt{1 + \frac{4r_{\rm e}^2}{d^2}} + \frac{1}
{\sqrt{1 + \frac{4r_{e}^2}{d^2}}}\right ).
\end{equation}

To evaluate the possibility of using particular approximations, it is
necessary to estimate the values of the task parameters. As has already
been mentioned, we consider only the white dwarf as a source of radiation.
The role of the lens is played by a white dwarf (WD), a neutron star (NS)
or a black hole (BH). At typical velocities of the orbital motion,
100--300 km/s, the accretion luminosity of the BH is $10^3 - 10^5$ as low as
the thermal luminosity of the WD (Ipser \& Price 1982). Assuming
$M_{WD} = 0.7 M_{\odot}, M_{NS} = 1.4 M_{\odot}, M_{BH} =
10 M_{\odot}, R_{NS} = 15 หอ, R_{BH} = r_g = 2GM_{BH}/c^2$, we have
estimated the rest of the parameters of binary systems, which are
presented in Table~1 in solar units.
\begin{table}[h]
\centering
\caption[]{Einstein radii for different binary systems}
\begin{tabular}{@{}c@{}ccccc@{}}
	    \hline
	    \noalign{\smallskip}
Pair&$M_l$      &$R_l$ &\multicolumn{2}{c}{$r_e$} & $R_s$\\
      &$(M_{\odot})$&$(R_{\odot})$&\multicolumn{2}{c}{($R_{\odot}$)}&$(R_{\odot})$\\
      &          &            &$a = 10$&$a = 100$        &  \\
\hline
\noalign{\smallskip}
WD+WD & 0.7 &$10^{-2}$       &$1.0\cdot 10^{-2}$&$3\cdot 10^{-2}$   &$10^{-2}$\\
WD+NS & 1.4 &$2\cdot 10^{-5}$&$1.1\cdot 10^{-2}$&$3.5\cdot 10^{-2}$&$10^{-2}$ \\
WD+BH & 10  &$4\cdot 10^{-5}$&$3\cdot 10^{-2}$  &$9.5\cdot 10^{-2}$&$10^{-2}$  \\
\noalign{\smallskip}
\hline
\end{tabular}
\end{table}

For the WD radius we used the relationship (Nauenberg 1972):
\begin{equation}
R =  0.01125 R_{\odot}\left (\left(\frac{M}{1.454 M_{\odot}}\right)^
{-\frac{2}{3}} -
\left(\frac{M}{1.454 M_{\odot}}\right)^{\frac{2}{3}}\right )^{\frac{1}{2}}.
\end{equation}

One can see that in a system of two WDs, the Einstein radius,
the sizes of the lens and the source are close. This implies that when
considering the light magnification one has to take into account the source
extension and occultation of its image by the companion -- the lens, at
least, at $a < 100 R_{\odot}$, which, as we will see, is our primary
interest. In binary systems, containing a NS and a BH, eclipsing
is absent; the source, however, can be regarded as a point one
only at $a > (10-20) R_{\odot}$, and one can apply expression (4) to derive
the brightness increase.

It should be noted that when the binary system is viewed
edge-on, at the moment of conjunction, a configuration in which the
observer, the lens and the source are on one line $(d = 0)$ is realized.
It was shown (Liebes 1964) that in this case $K = \sqrt{1 + \frac{4r_{e}^2}
{R_{s}^2}}$, as a result of averaging of expression (4) over the surface
of the uniformly radiating disk. At $d\ne 0$ the task gets essentially
more complicated and has no analytical solution. The accurate relationships
for the brightness magnification by a point lens, where elliptical
Legendre integrals of the first and second kinds are used, were derived by
Refsdal (1964), Liebes (1964), Byalko (1969), Ingel (1972, 1974) and
Maeder (1973). There is, however, one more case in which a
simple expression for $K$ exists -- the contact between the disk edge
and the lens centre:
\begin{equation}
\begin{array}{ll}
K &= \frac{2}{\pi}\left [1+4(\frac{r_e}{R_s})^2\right ]\arcsin
\left [1+4(\frac{r_e}{R_s})^2\right ]^{-\frac{1}{2}} \\
&\\
&\approx \frac{2}{\pi}\sqrt{1+4\beta^2}\, ,
\end{array}
\end{equation}
where $\beta = \frac{r_e}{R_s}$.\\
\begin{figure}[h]
\centering
\includegraphics[width=8.5cm,height=7.3cm]{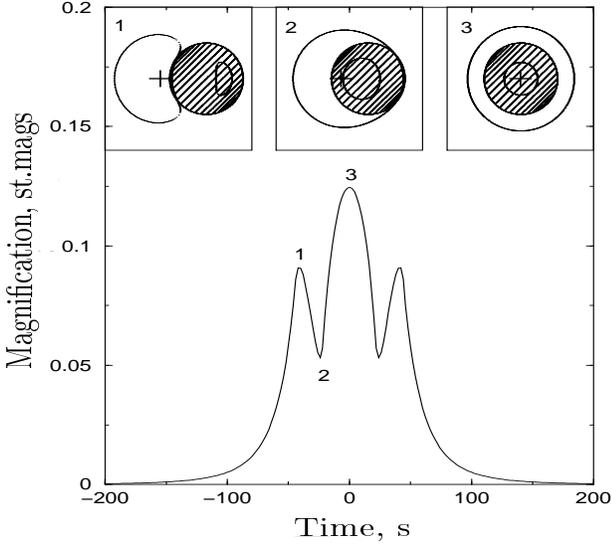}
\caption{Magnification curve for the pair of white dwarfs.
The crosshatched circle is the gravitational lensing component.}
\end{figure}

The light curve of the component-source is the result of variation of
magnification with time variations of $d$ during the system rotation period.
In the approximation of a point source, according to (4)
\begin{equation}
K(t) = \frac{1}{2}\left [\sqrt{1 + \frac{4r_e^2}{d(t)^2}} + \frac{1}
{\sqrt{1+\frac{4r_e^2}{d(t)^2}}} \right ].
\end{equation}
Near the conjunction $d^2(t) = d_0^2 + (vt)^2$, where $d_0 = a\sin i$ is
the minimum value of $d$, which depends on the orbit inclination
$i$, $v$ is the ralative velocity of the system components.
At the moment $t=0$, the brightness variation reaches a maximum
and
\begin{equation}
\begin{array}{ll}
K_m&=K(0) = \frac{1}{2}\left [\sqrt{1+\frac{4r_e^2}{d_0^2}} + \frac{1}
{\sqrt{1+\frac{4r_e^2}{d_0^2}}}\right ] \\
&\\
&=\frac{1}{2} \left [\sqrt{1+4\beta_{0}^2} +
\frac{1}{\sqrt{1 + 4\beta_{0}^2}}\right ],
\end{array}
\end{equation}
here $\beta_0 = \frac{r_e}{d_0}$.

It is customary to characterize the duration of the flare (brightness
increase) (see Zakharov \& Sazhin 1998) by the time $\tau$ that is taken
by the moving source projection onto the lens plane to cross the Eintstein
ring. It can be easily found from the relation $d^2(\frac{\tau}{2}) =
d_0^2 + (\frac{v\tau}{2})^2 = r_e^2$, that:
\begin{equation}
\tau = \frac{2r_e}{v}\sqrt{1 - \frac{1}{\beta_0^2}}.
\end{equation}
In a point approximation
$K(\frac{\tau}{2}) = \frac{1}{2}[\sqrt{5} + \frac{1}{\sqrt{5}}]\doteq 1.34$,
and $\tau$ defines the flare duration at a level $\Delta m_0 = 0\fm 32$
of the source light at a quiescent phase.

Note that the brightness magnification of one of the
components by a factor of $K$ corresponds to magnification of the system
brightness (which is the subject of detection in observations) $K_0$
times, and
\begin{equation}
K = K_0 + \frac{L_l}{L_s}(K_0 - 1),
\end{equation}
where $\frac{L_l}{L_s}$ is the lens--source brightness ratio. For a
system containing two equal white dwarfs $\frac{L_l}{L_s}= 1$,
$K_0 = \frac{1}{2}(K+1)$, and $\Delta m_0 = 0\fm 17$ with respect to
the total system brightness off the flare. It is self-evident that if the
part of the lens is played by a neutron star or a black hole, the
$K$ is practically equal to $K_0$.

Present-day telescopes detect rather small light variations during short
exposures (see Section~5).
We are guided by these possibilities when determining the minimum
amplitude of detectable flares. Say, if the amplitude of a flare
$\Delta m_{0} = 0\fm 2$ ($K = 1.2$), then the trajectory of the source
projection onto the lens plane does not cross the Eintstein ring at all,
and such a flare is excluded from the analysis. Therefore, it is necessary
to introduce $\tau_{0.1}$ at which $K = 1.1$. Using (4) we derive
\begin{equation}
(\frac{r_e}{d})^2 = \beta^2 = \frac{K^2 - 1 + K\sqrt{K^2 - 1}}{2},
\end{equation}
hence at $K = 1.1, \beta_{0.1} = 0.60$, ษ $d_{0.1} = 1.67 r_e$.
Thus
\begin{equation}
\tau_{0.1} = 1.67\frac{2r_e}{v}\sqrt{1 - \frac{1}{(1.67\beta_0)^2}}.
\end{equation}

In the general case, when deriving the light curves, equation (1) was solved
for each point of the source profile, its image was constructed in the lens
plane, and its area was found numerically. Since in gravitational
lensing, the surface brightness remains constant, the light curve is
determined as variation of the ratio of the image areas and the source
itself, depending on the position of the binary system components with
respect to the observer, which varies during the period of rotation.
The contribution to the system radiation of the lens--white dwarf
and occultation by it of a part of the source image was taken into account.
Fig.~2 shows the brightness variations of the
binary system consisting of two white dwarfs with a mass of $0.7 M_{\odot}$,
$a = 14.2 R_{\odot}$ and the orbit inclination $i = 0$ (i.e. $d_0 = 0$).
The similarity of the sizes of the source, the lens and the Einstein ring
results in formation of a rather complex light curve.
Similar events were analysed by Marsh (2001) and Agol (2002).

Comparison of the light curves in Fig.~3, plotted from the use of the
results of direct calculations and with the use of point approximation (7),
shows that they are virtually coincident for  $a > 30R_{\odot}$.
In this case the light curves can be constructed by means of a point
approximation. The exception is the region near the maximum at
$\frac{d(t)}{r_e} < 0.25$.
\begin{figure}[h]
\centering\includegraphics[width=8.5cm]{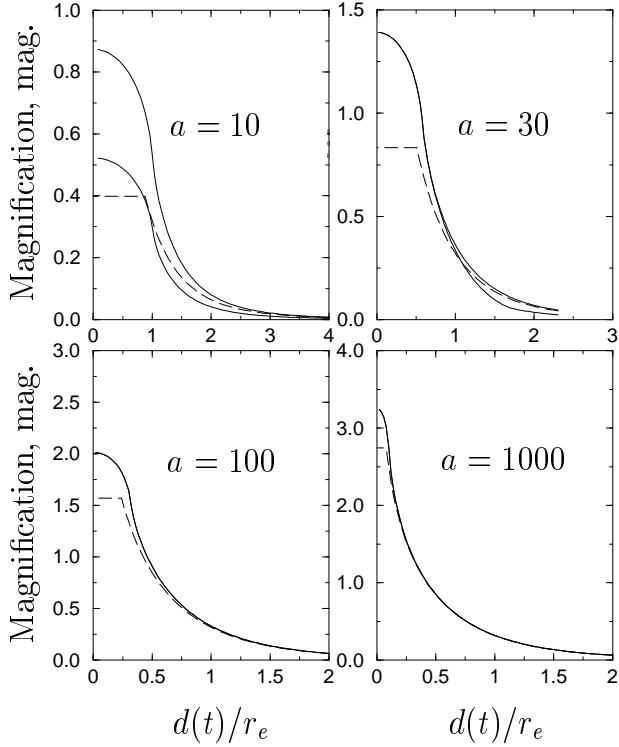}
\caption{Magnification curves of the pair of white dwarfs for different
semiaxes $a$. Dashed lines represent the analytical approximation (7).
Solid lines for numerical calculation; fine ones for transparent lens,
thick ones for opaque lens.}
\end{figure}

Consider the effect of gravitational lensing in a binary system comprising
two equal white dwarfs with $M_{WD} = 0.5 M_{\odot}$ and $a = 100 R_{\odot}$.
According to Kepler's third law (in solar units):
\begin{equation}
T = 2.79a^{3/2}(M_l + M_s)^{-\frac{1}{2}}\; {\rm hours},
\end{equation}
its period $T = 82.4$ days.
Flares which amplitude depends on $i$ will be observed twice during
this time. For $d_0 \approx ai > R_s$,
point approximation (8) can be employed to estimate the amplitude.
Consequently, at $i > \frac{R_s}{a} = \frac{10^{-2}}{10^2} = 10^{-4}$,
it follows from (2) and (8)
$$
K_m = \frac{1}{2}\left[\sqrt{1+\frac{4}{i^2}\cdot \frac{2r_g}{a}} +
\frac{1}{\sqrt{1+\frac{4}{i^2}\frac{2r_g}{a}}}\right ]\le 2.27.
$$
Thus the amplitude of these flares $K_0 = \frac{1}{2}(K_m + 1)\le 1.64$
or $\Delta m_0\le 0.54$.

From (2) and (9), considering that $v = \sqrt{\frac{G(M_l+M_s)}{a}}$,
obtain under the same condition (point approximation):
\begin{equation}
\tau = \frac{4a}{c}\sqrt{\frac{M_l}{M_l + M_s}}\cdot \sqrt{1 - \frac{i^2 a}
{9\cdot 10^{-6}M_l}}.
\end{equation}
And for the given system $\tau = \frac{4a}{c\sqrt{2}}\sqrt{1 - \frac{i^{2}a}
{4.5\cdot 10^{-6}}} = 662\sqrt{1 - 2.2\cdot 10^7\cdot i^2}$ sec.

\begin{figure}[h]
\centering\includegraphics[width=8.5cm]{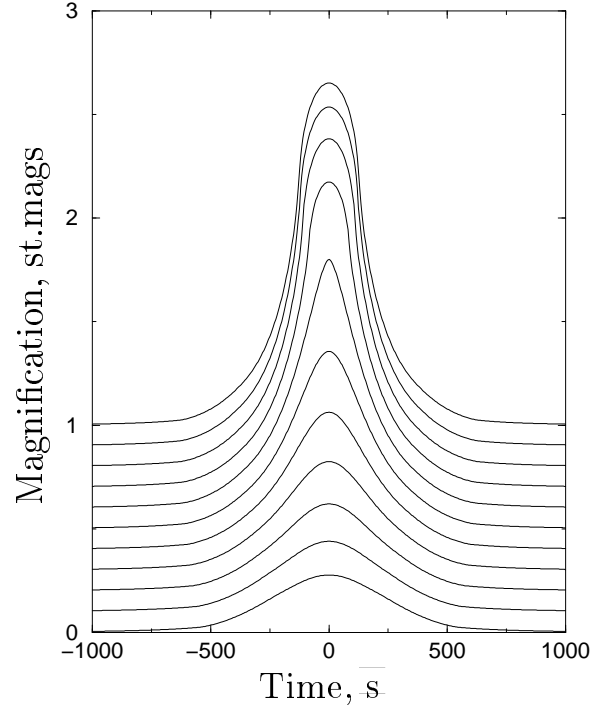}
\caption{Simulated light curves of the pair of white dwarfs.}
\end{figure}

In Fig.~4 the light curves of this system for different $i$,
plotted from the data of numerical computations, are displayed.
The light variations are expressed in stellar magnitudes.
The orbit inclination $i$ varies from zero to
$2.5\cdot 10^{-4}$ with a step of $2.5\cdot 10^{-5}$ from the upper to
the lower curve, which is displaced for convenience. The flare shape
variations are due to these changes.
\begin{table}[h]
\centering
\caption[]{Parameters of the closest pairs able to give rise to a flare
of amplitude $\Delta m$}
\begin{tabular}{@{}cccccccc@{}}
	    \hline
	    \noalign{\smallskip}
$\Delta m$&\multicolumn{3}{c}{WD+WD}     &\multicolumn{3}{c}{WD+NS}     &WD+BH\\
 &$a_{min}$&$a_{c}$&$\tau_{0.1}$&$a_{min}$&$a_{c}$&$\tau_{0.1}$&$a_{min}$  \\
 &($R_{\odot}$)&($R_{\odot}$)&(s)&($R_{\odot}$)&($R_{\odot}$)&(s)&($R_{\odot}$)\\
\hline
\noalign{\smallskip}
$0\fm 2$ & 4 &14&44  &0.92&2.5&12&0.13\\
$0\fm 5$ &15.4&23&170&3.15&3.2&41&0.44 \\
$1\fm 0$ &62.4&49&690&11  &10 &141&1.52  \\
\noalign{\smallskip}
\hline
\end{tabular}
\end{table}

One more step brings us closer to the real condition of
searching for flares. Having fixed three levels of brightness increase,
$0\fm 2$, $0\fm 5$ and $1\fm 0$, we determined by two methods the
minimum sizes of the systems consisting of two white dwarfs and also a
white dwarf and a neutron star whose light rises to these levels (Table~2).
From the relation $K = \sqrt{1 + \frac{4r_e^2}{R_s^2}}$, $a_{min}$ has been
found analytically, while $a_c$ has been derived by direct calculations.
The light curves of flares in these edge-on-viewed pairs,
which were constructed based on numerical computations, are presented
in Fig.~5a, b. It can be seen from Table~2 and this figure that the
analytical estimates of the characteristic durations of flares
$\tau_{0.1}$ are close to the results of direct calculations. The flares
of a white dwarf coupled with a black hole  constructed in a similar
manner are exhibited in Fig.~5c (Table~2 contains only $a_{min}$ for
this binary system), here the separation of the components $a = 3R_{\odot}$,
while the amplitudes $0\fm 2, 0\fm 5, 1\fm 0$ correspond to the orbital
plane inclinations 0.023, 0.015 and 0.01. It is precisely this value of
$a$ that was used in the computation since it is close to the minimum
size of the pair WD--BH, which "survived" in the process of emission of
gravitation waves during the lifetime of the Galaxy. The number of closer
pairs drops sharply and the probability of their detection is very low
(see Section~3).
\begin{figure}[h]
\centering\includegraphics[width=8.5cm,
]{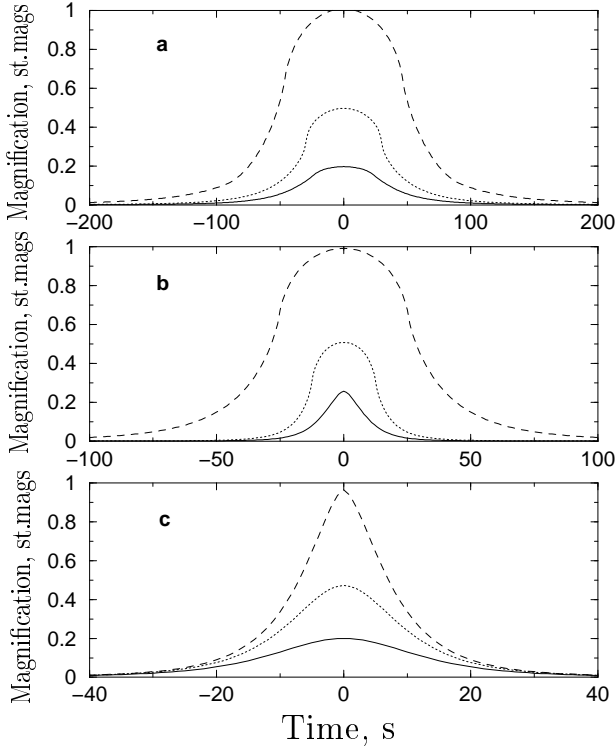}
\caption{Simulated light curves of the flares as a result of gravitational
lensing in different pairs: a)~ WD--WD, b)~ WD--NS, c)~ WD--BH. Dashed
lines show flares with amplitude $1\fm 0$, dotted lines are used for flares
with amplitude $0\fm 5$ and solid lines for flares with amplitude $0\fm 2$.}
\end{figure}

Fig.~5a, b, c shows that the characteristic durations of flares for
pairs of white dwarfs are 50--200 s; for binaries containing NS and BH,
the light variations are shorter, 30--130 s and 20--40 s, respectively.
Singnificant variations of the flare shape are noticeable in all the cases.

Without dwelling on details, note that the set of parameters of these
binary systems obtained in observations -- light curves, periods, amplitudes,
duration and shape of flares, can make it possible to determine physical
characteristics of the components -- their masses, sizes, temperatures
(e.g. Cherepashchuk \& Bogdanov 1995a, b). This problem can be resolved
completely for two white dwarfs. At the same time, the determination of,
at least, the mass of the lens component and, therefore, its identification
as a NS or a BH seems to be self-important for the study of properties of
relativistic objects and their extreme gravitation fields.

\section{Probability of detection of flares and their average
characteristics}

The number of systems that can be found as a result of brightness increases
in gravitation lensing is determined, first of all, by the probability
$F(K_{0},t)$ of recording a flare with an amplitude larger than the level
$K_0$, during the observing time $t$ of a sample of systems
distributed in some manner over parameters (orbital separation, masses and
component brightness).
This means that $F(K_{0},t)$ is the probability of a
flare detection from a certain system averaged over such a distribution
\begin{equation}
F(K_{0},t) = \int\limits_{\Omega} P(K_{0}, t, \omega) f(\omega )d\omega,
\end{equation}
where $\omega$ is the set of parameters, $f(\omega)$ is the function of
their distribution, $\Omega$ is the region in which they are specified,
$P(K_0, t, \omega)$ is the probability of recording of the flare in a
system with the parameters $\omega $. For the sake of simplicity, with
allowance made for the low accuracy of any estimates associated with
statistical characteristics of binary systems with compact companions
(Masevich \& Tutukov 1988; Lipunov et al. 1996; Nelemans et al. 2001;
Fryer et al. 1999, where the errors of different models are defined),
we restrict ourselves to an analysis of the distribution of the systems
only in $a$. Then $P(K_{0}, t, a)$ is the probability of a joint onset of
two events, favourable orientation of the orbital plane with the semiaxis
$a$, when the level $K_0$ in the flare is exceeded, and the recording of
a result if only one such a flare occurs during the time $t$,
$f(\omega)d\omega = f(a)da$.

The first event is consistent with the geometric probability $p_g$
which can be easily found. It follows from (4), (7), (8) and (11) that
the values $d_0^{\prime} < d_0 =\frac{r_e}{\beta(K)}$,
where $K = K_0 +\frac{L_l}{L_s}(K_0 - 1)$, correspond to the flares of the
source with amplitudes exceeding $K$. The locus of the ends of the
normals of the orbital planes satisfying this condition is a globe belt of
width $d_0$ and radius $a/2$, its area is $S_d = \pi ad_0$. The locus of the
ends of the normals of any orbit is the surface of a sphere of radius
$\frac{a}{2}$, its area is $S_a = \pi a^2 $. Hence $p_g = \frac{S_d}{S_a} =
\frac{d_0}{a}= \frac{r_e}{a\beta (K)}$. Finally, from~(2) 
$p_g = 3\cdot 10^{-3} a^{-\frac{1}{2}} M_l^{\frac{1}{2}}[\beta (K)]^{-1}$.

Note that in determining $\beta (K)$, both analytical expression
(11) and the numerical relationship can be applied.
Fig.~6 shows the relationships between geometric probabilities and the size
of a system consisting of two white dwarfs which have been derived by
these two methods (analytically in a point approximation -- the dashed
line, and numerically -- the solid one). The contribution of the dwarf
lens is taken into account here. Minimum sizes of systems capable of
giving rise to a flare of amplitude $K$ correspond to the values
$a_c$ in Table~2. The numerical calculations in systems WD--NS and WD--BH
are virtually the same as the analytical estimates, but $p_g$ for WD-NS
reduces to zero at $a = a_c$.
\begin{figure}[h]
\centering\includegraphics[width=8.5cm,]{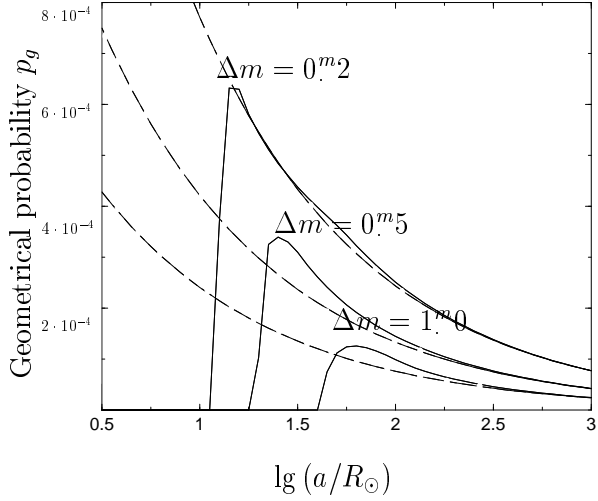}
\caption{Geometrical probability versus semiaxis size for the pair of white
dwarfs. Dashed lines for analytical approximation, solid lines for
numerical calculation.}
\end{figure}

The probability $p_t$ of recording a flare during the time $t$ is
essentially dependent on the relationship of the period of recurrence
of flares $T$ (given by~(13)), the duration of the
flare $\tau$ and the net continuous time of observing $t$.
Let us lay down certain conditions which must be satisfied to make the
search for flares efficient enough. First of all, it is evident that the
duration of an elementary continuous exposure $t_e$ must be close to
that of flare $\tau$. On the other hand, the interval between the exposures
must be considerably shorter than $\tau$. Implying by $t = \sum\limits_e
t_e$ the duration of the total exposure when observing a specified sky
region, we obtain for the probability $p_t$:
\begin{equation}
p_t = \left\{
\begin{array}{ccc}
\frac{t}{T}&{\rm at} &t < T\\
1 &{\rm at}& t > T,\\
\end{array}
\right.
\end{equation}

which corresponds to
$$
p_t =\left\{
\begin{array}{ccl}
0.36\frac{t(M_l + M_s)^{1/2}}{a^{3/2}}&{\rm at}&a > a_t  \\
1 &{\rm at}& a < a_t,
\hspace*{2.7cm}(16^{\prime})\\
\end{array}
\right.
$$
where $a_t = [0.36t(M_l + M_s)^{1/2}]^{2/3}$ is the semiaxis of binary 
with period $T=t$ and we consider $a_t\geq a_c$ only since $p_g=0$ as $a<a_c$.

Hence

\begin{equation}
\begin{array}{l}
P(K_{0}, t, a) = p_g p_t \\
\; =\left\{
\begin{array}{lll}
0,&{\rm at} &a < a_c,\\
3\cdot 10^{-3}a^{-\frac{1}{2}}M_l^{1/2}[\beta (K)]^{-1}&{\rm at}& a_c < a
< a_t,\\
&&\\
1.08\cdot 10^{-3}ta^{-2}
\frac{[M_l(M_l + M_s)]^{1/2}}
{\beta(K)}&{\rm at}& a_c <a;\\
&&a_t < a .\\
\end{array}
\right.
\end{array}
\end{equation}

Due to the serious anisotropy of pair distributions on the sky, the optimum
time of observation of a specified sky field will be shown to be greater than 
one night length (at least for WD-WD systems). In this case
a simple expression (16) for $p_t$ no longer holds since observations on 
individual nights become statisticaly independent.
The probability of detecting at least one flare in $n$ nights is given
by the Bernoulli distribution:
$$
	p_n=1 - (1-p_1)^n=\sum\limits_{1}^n (-1)^{i+1}C_n^i p_1^i. 
$$
  	where $p_1=p_t|_{t=9^h}=3.24(M_l+M_s)^{1/2}a^{-3/2}$ is 
the detection probability in $9^h$ (the observational night average length).

	Thus, expression for $P(K_0,n,a)$ takes the following form
$$
\begin{array}{l}
P(K_{0}, n, a) = p_g p_n \\
\; =\left\{
\begin{array}{lll}
0,&{\rm at} &a < a_c\\
3\cdot 10^{-3}a^{-\frac{1}{2}}M_l^{1/2}[\beta (K)]^{-1},&{\rm at}& a_c < a
< a_1 \\
3\cdot 10^{-3}M_l^{1/2}[\beta(K)]^{-1} & & \\
\times\sum\limits_{1}^n(-1)^{i+1}(3.24)^iC_n^i &&\\
\times (M_l+M_s)^{i/2} a^{-\frac{1}{2}(3i+1)} &{\rm at}& a_1 < a,
\hspace*{1.3cm}(17^{\prime})
\end{array}
\right.
\end{array}
$$
where $a_1=a_t|_{t=9^h}=2.2(M_l+M_s)^{1/3}$.

 For randomization corresponding to~(15) we must convolve $P(K_0,t,a)$
or $P(K_0,n,a)$ with the pair distribution function in $a$.

Now we turn to the analysis of this distribution.
We must clearly see the difference in distribution for binary systems,
progenitors of pairs of compact objects, and the finite distribution
of these pairs themselves. In the first case the results of observations yield
a distribution uniform in logarithm of semiaxis (Abt 1983):
$f (a)\propto a^{-1} (1\le \log(a/R_{\odot})\le 6)$. In the second case
the situation is not so clear because of the uncertainty in quantitative
characteristics determined in the course of different versions of population
synthesis and because of the qualitative distinctions of these
versions (Yungelson et al. 1994; Lipunov et al. 1996; Saffer et al. 1998;
Portegies Zwart \& Yungelson 1998; Fryer et al. 1999; Nelemans et al. 2001).
For instance, in different papers the initial mass functions with different
indices are used; due to the great uncertainty in the loss of matter, the
supposed masses of BH progenitors vary from $25 M_{\odot}$ to $80 M_{\odot}$.
The finite numbers of, say, couples WD--BH contain an uncertainty of 5
orders of magnitude (see Fryer et al. 1999)! Nevertheless, since our
analysis is qualitative, one can be content with the main
characteristics of samples of objects and rough estimates.

First of all, it will be noted that couples lose energy and angular momentum
radiating gravitation waves (Landau \& Livshits 1983; Grishchuk et al. 2001).
Since the rate of these losses depends dramatically on  $a\;(\propto a^{-5})$,
the stars in close pairs stick together and leave the sample. Using the
expression for gravitational luminosity of a binary system in a circular
orbit $\frac{dE}{dt} = -\frac{32}{5}\frac{G^4}{c^5}\cdot\frac{M_l^2 M_s^2
(M_l + M_s)}{a^5}$ (Landau \& Livshits 1983) and taking into account
that the total energy of the system $E = -\frac{GM_l M_s}{2a}$, one can
readily derive the value of $a_r$ of the initial semiaxis of a system
that will merge during $10^{10}$ years, which is the life-time of the Galaxy:
$$
a_{r} = 2 R_{\odot}[M_l M_s (M_l + M_s)]^{1/4}.
$$

We tabulate in Table~3 the values of $a_r$ and corresponding $T_r$ for
the systems under study.
\begin{table}[h]
\centering
\caption[]{The closest pairs that survived during the life-time of the
Galaxy}
\vspace*{0.3cm}
\begin{tabular}{cccc}
	    \hline
	    \noalign{\smallskip}
Pair      &$M_l + M_s$     &$a_r$        &$T_r$\\
	  &($M_{\odot}$)   &($R_{\odot}$)&{\rm (hours)}  \\
\hline
\noalign{\smallskip}
WD+WD &0.7+0.7   &1.82  &5.77\\
WD+NS &0.7+1.4   &2.40  &7.11 \\
WD+BH &0.7+10    &5.88  &12.13  \\
\noalign{\smallskip}
\hline
\end{tabular}
\end{table}

The results of population synthesis, which are partially confirmed by
observations, suggest that the density destribution of pairs of compact
objects rises with decreasing $a$ approximately as $\frac{1}{a}$,
reaches a maximum at $a = a_0\sim a_r$, after which it decreases rapidly
to $a\sim 1R_{\odot}$ (Lipunov et al. 1996; Hernanz et al.
1997; Saffer et al. 1998; Portegies Zwart \& Yungelson 1998; Wellstein
\& Langer 1999).

It follows from the same papers that another important parameter, the
maximum size of a binary system $a_{max}$, is known to be less than
$50 R_{\odot}$ at small eccentricities.

With allowance made for the said above, we have used the function of
distribution in $a$ in a simple form:
\begin{equation}
\begin{array}{l}
f(a) = \gamma\lg eg(a) \\
\qquad {}=\left\{
\begin{array}{lcl}
\frac{\gamma \lg e}{a_0(a_0 - 1)}(a - 1), &{\rm at} & 1\le a\le  a_{0},\\
&&\\
\frac{\gamma \lg e}{a},&{\rm at} & a_0\le a\le  a_{max}.\\
\end{array}
\right.
\end{array}
\end{equation}
From the condition of normalizing $\int\limits_1^{a_{max}}f(a)da = 1$ one can
easily find $\gamma$:
\begin{equation}
0.43\gamma = \frac{1}{\frac{a_0 - 1}{2a_0} + \ln \frac{a_{max}}{a_0}}.
\end{equation}
Note at once that the share of systems with $a < a_0$ ranges from
10~\% to 30~\% at $3 < a_0 < 10$ and $a_{max} = 50, 30$, which is close
to the results of population synthesis for pairs of different type compact
objects (Hernanz et al. 1997; Saffer et al. 1998; Portegiez Zwart \&
Yungelson 1998; Nelemans et al. 2001).

In our further estimations we will use distribution (18), varying the
parameter $a_0$ in the range $3R_{\odot}\div 10R_{\odot}$, at two
values of the parameter $a_{max} - 30R_{\odot}$ and $50R_{\odot}$.

Having fixed the distribution $f(a)$, now we have to determine the lower
limits of integrating in (15) for different types of binary systems.
Referring to Table~2, Table~3 and Fig.~6 and allowing for (17) and (18),
it can be readily seen that the lower limits for systems consisting of
two white dwarfs and a white dwarf in pair with a neutron star coincide
with $a_c$  and are 14 and 2.5, respectively. However, for pairs
consisting of a white dwarf and a black hole integration is done over the
whole range of possible values of $a$, i.e. from 1 to $a_{max}$.
The values of binary periods corresponding to $a_c$ are 123, 7.6 and 0.85
hours for WD-WD, WD-NS and WD-BH systems respectively.
This leads to the following expressions for flare detection probability 
obtained from (16), ($16^{\prime}$), (17), ($17^{\prime}$) and (18).

For WD-WD pairs\\
when $t<9^h<T_c=123^h$ since $a|_{t=9^h}<a_c$ and $a_0<a_c$,
\begin{equation}
\begin{array}{l}
F(K_0,t)
= \int\limits_{a_c}^{a_{max}} f(a) 1.08\cdot 10^{-3} [\beta(K)]^{-1} t a^{-2}\\
\quad\times [M_l(M_l+M_s)]^{1/2} da
=1.08\cdot 10^{-3}\cdot 0.43\gamma \\
\quad\times[\beta(K)]^{-1} t [M_l(M_l+M_s)]^{1/2}
 \int\limits_{a_c}^{a_{max}} a^{-3} da \\
\qquad\qquad\:=5.4\cdot 10^{-4}[\beta(K)]^{-1}
t[M_l(M_l+M_s)]^{1/2} \\
\quad\times\left[\frac{a_0-1}{2a_0}+\ln\frac{a_{max}}{a_0}\right]^{-1}
\left(\frac{1}{a^2_c}-\frac{1}{a^2_{max}}\right) ;
\end{array}
\end{equation}
when the overall observation time of a specified sky region is $n$ nights,

$$
\begin{array}{l}
	F(K_0,n) = \int\limits_{a_c}^{a_{max}} f(a)3\cdot 10^{-3} M_l^{1/2}
[\beta(K)]^{-1} \\
\quad\times\sum\limits_{i=1}^n(-1)^{i+1} (3.24)^i C_n^i (M_l+M_s)^{\frac{i}{2}}
a^{-\frac{3i+1}{2}} da \\
\qquad\qquad\:= 3\cdot 10^{-3} 0.43\gamma M_l^{1/2}[\beta(K)]^{-1} \\
\quad\times \sum\limits_{i=1}^n(-1)^{i+1}(3.24)^iC_n^i(M_l+M_s)^{\frac{i}{2}}
\int\limits_{a_c}^{a_{max}} a^{-\frac{3(i+1)}{2}} da \\
\qquad\qquad\:= 3\cdot 10^{-3}\left[\beta(K)\left(\frac{a_0-1}{2a_0}+\ln
\frac{a_{max}}{a_0}\right)\right]^{-1}\\
\quad\times M_l^{1/2}\sum\limits_{i=1}^n(-1)^{i+1}(3.24)^iC_n^i (M_l+M_s)^
{\frac{i}{2}}\frac{2}{3i+1}\\
\quad\times\left[a_c^{-\frac{3i+1}{2}} - a_{max}^{-\frac{3i+1}{2}} \right].
\hspace*{4.4cm}(20^{\prime})
\end{array}
$$
For WD-NS pairs\\
when $t<T_c=7.^h 6$, since $a_t<a_c<a_0$
\begin{equation}
\begin{array}{l}
F(K_0,t) = 1.08\cdot 10^{-3}[\beta(K)]^{-1} t [M_l(M_l+M_s)]^{1/2}\\
\quad\times\int\limits_{a_c}^{a_{max}} f(a) a^{-2} da
= 1.08\cdot 10^{-3}\\
\quad\times\left[\beta(K)\left(\frac{a_0-1}{2a_0}+\ln\frac{a_{max}}
{a_0}\right)\right]^{-1}
t [M_l(M_l+M_s)]^{1/2} \\
\quad\times\Bigl[ \frac{1}{a_0(a_0-1)}\left[\ln\frac{a_0}{a_c}-\left(\frac{1}
{a_c}-\frac{1}{a_0}\right)\right]+\frac{1}{2}\left( \frac{1}{a^2_c}-\frac{1}
{a^2_{max}}\right)
\Bigr];
\end{array}
\end{equation}
	when $7.^h 6 <t<9^h$, since $a_c<a_t<a_0$

$$
\begin{array}{l}
F(K_0,t)=3\cdot 10^{-3} [\beta(K)]^{-1} M_l^{1/2}
\biggl [\int\limits_{a_c}^{a_t}a^{-1/2}f(a)da   \\
\quad + 0.36(M_l+M_s)^{1/2}t\int\limits_{a_t}^{a_{max}}a^{-2}f(a)da
\biggr ] \\
\qquad\qquad = 3\cdot 10^{-3} \biggl[\beta(K)\left(\frac{a_0-1}{2a_0}+\ln
\frac{a_{max}}{a_0}\right )\biggr ]^{-1} M_l^{1/2} \\
\quad\times\biggl \{\frac{\frac{2}{3}[0.36t(M_l+M_s)^{1/2}-a_c^{3/2}]
-2[0.71t^{1/3}(M_l+M_s)^{1/6}-a_c^{1/2}]}{a_0(a_0-1)} \\
\quad +0.36t(M_l+M_s)^{1/2}
\biggl [\frac{1}{a_0(a_0-1)}
\biggl [\ln a_0 -\frac{2}{3}\ln \biggl(0.36t \\
\quad\times (M_l+M_s)^{1/2}\biggr )
-  \biggl (0.36t(M_l+M_s)^{1/2}\biggr )^{-\frac{2}{3}}+\frac{1}{a_0}\biggr ]\\
\quad +
\frac{1}{2}\left(\frac{1}{a^2_0}-\frac{1}{a^2_{max}}\right)\biggr ]
\biggr \}; \hspace*{4.5cm}(21a)
\end{array}
$$
when the overall observation time of a specified sky region is $n$ nights,
since $a_c=2.5<a_t|_{t=9^h}=2.8$ and $a_t<a_0$
$$
\begin{array}{l}
F(K_0,n) = 3\cdot 10^{-3}[\beta(K)]^{-1} M_l^{1/2}\biggl[\int\limits_{a_c}^{a_t}
f(a)a^{-1/2}da \\
\;\; +\sum\limits_{i=1}^n(-1)^{i+1}(3.24)^iC_n^i (M_l+M_s)^{\frac{i}{2}}
\int\limits_{a_t}^{a_{max}} f(a) a^{-\frac{3i+1}{2}} da \biggr]\\
\qquad\qquad\; =3\cdot 10^{-3}\biggl[\beta(K)\left(\frac{a_0-1}{2a_0}+\ln\frac
{a_{max}}{a_0}
\right) \\
\;\;\times (a_0(a_0-1))\biggr]^{-1}
 M_l^{1/2}
\biggl[\frac{2}{3}[3.24(M_l+M_s)^{1/2}-a_c^{3/2}]\\
\;\; -2[1.48(M_l+M_s)^{1/6} -a_c^{1/2}]+ 3.24n(M_l+M_s)^{1/2} \\
\;\;\times \ln\frac{a_0}{2.2(M_l+M_s)^{1/3}}
+\sum\limits_{i=2}^n(-1)^{i+1}(3.24)^iC_n^i(M_l+M_s)^{\frac{i}{2}} \\
\;\;\times \frac{2}{3(i-1)}\left[(3.24(M_l+M_s)^{1/2})^{-(i-1)} -
a_0^{-\frac{3(i-1)}{2}} \right]  \\
\;\; -\sum\limits_{i=1}^n(-1)^{i+1}(3.24)^iC_n^i(M_l+M_s)^{\frac{i}{2}}\\
\;\;\times\biggl[\frac{2}{3i-1}\biggl[(3.24(M_l+M_s)^{1/2})^{-(i-
\frac{1}{3})}-a_0^ {-\frac{3i-1}{2}}\biggr]\\
\;\; -\frac{2}{3i+1}(a_0^{-\frac{3i+1}{2}}-a_{max}^{-\frac{3i+1}{2}}) \biggr]
\biggr]. \hspace*{3.6cm}(21^{\prime})
\end{array}
$$
For WD-BH pairs    \\
when $t<T_c=0.^h 85$, since $a_t<a_c<a$ and $a_c<a_0$ where $a_c=1$ (left
distribution edge)
\begin{equation}
\begin{array}{l}
F(K_0,t)=1.08\cdot 10^{-3}[\beta(K)]^{-1} t[M_l(M_l+M_s)]^{1/2}\\
\quad\times\int\limits_1^{a_{max}} f(a)a^{-2} da =
1.08\cdot 10^{-3}\biggl[\beta(K)\biggl(\frac{a_0-1}{2a_0}\\
\quad +\ln\frac{a_{max}}{a_0}\biggr)\biggr]^{-1} t [M_l(M_l+M_s)]^{1/2}\\
\quad\times\biggl[\frac{1}{a_0(a_0-1)}\left(\ln{a_0}-1+\frac{1}{a_0}\right)+
\frac{1}{2}\left(\frac{1}{a_0^2} -\frac{1}{a^2_{max}}\right)\biggr];
\end{array}
\end{equation}
	when $0.^h 85<t<9^h$, since $1<a_t<a_t|_{t=9^h}=4.8$
$$
\begin{array}{ll}
F(K_0,t)=&3\cdot 10^{-3}[\beta(K)]^{-1} M_l^{1/2}
\biggl[\int\limits_1^{a_t}a^{-1/2}f(a)da \\
&+ 0.36(M_l+M_s)^{1/2} t \int\limits_{a_t}^{a_{max}}a^{-2} f(a) da\biggr] ,
\end{array}
$$
	and for $a_t<a_0$
$$
\begin{array}{l}
F(K_0,t)=3\cdot 10^{-3}\left[\beta(K)\left(\frac{a_0-1}{2a_0}+\ln
\frac{a_{max}}{a_0}\right)\right]^{-1} \\
\quad\times M_l^{1/2}
\biggl[\frac{0.24t(M_l+M_s)^{1/2}-1.42t^{1/3}(M_l+M_s)^{1/6}+\frac{4}{3}}
{a_0(a_0-1)}\\
\quad +0.36t(M_l+M_s)^{1/2}
\biggl[\frac{1}{a_0(a_0-1)}\biggl[\ln{a_0}\\
\quad -\frac{2}{3}\ln{[0.36t(M_l+M_s)^{1/2}]}
- 2.8t^{-2/3}(M_l+M_s)^{-1/3}\biggr]                \\
\quad +\frac{1}{2}(a^{-2}_0-a^{-2}_{max})\biggr]\biggr],\hspace*{4.5cm}(22a)
\end{array}
$$
	or for $a_0<a_t$

$$
\begin{array}{l}
F(K_0,t)=3\cdot 10^{-3}\left[\beta(K)\left(\frac{a_0-1}{2a_0}+\ln
\frac{a_{max}}{a_0}\right)\right]^{-1} \\
\quad\times M_l^{1/2}\Bigl \{\frac{1}{a_0(a_0-1)}(\frac{2}{3}a_0^{3/2} -
2a_0^{1/2}+\frac{4}{3})+2\biggl[a_0^{-1/2}\\
\quad -1.41t^{-\frac{1}{3}}(M_l+M_s)^{-\frac{1}{6}} \biggr]+0.18
(M_l+M_s)^{1/2}t\\
\quad\times\left[3.8t^{-\frac{4}{3}}(M_l+M_s)^{-\frac{2}{3}}
- a^{-2}_{max}\right]
\Bigr\} ; \hspace*{2.3cm}(22b)
\end{array}
$$
when the overall observation time of a specified sky region is $n$ nights,
for $a_1=2.2(M_l+M_s)^{1/3}=4.8<a_0$, $F(K_0,n)$ coincides with
($21^{\prime}$) where $a_c$ is replaced by 1, and for $a_0\le 4.8$
$$
\begin{array}{l}
F(K_0,n)=3\cdot 10^{-3} M_l^{1/2}[\beta(K)]^{-1}\Bigl(\int\limits_1^{a_1}f(a)
a^{-1/2}da\\
\quad +\sum\limits_{i=1}^n(-1)^{i+1}(3.24)^i(M_l+M_s)^{\frac{i}{2}}C_n^i
\int\limits_{a_1}^{a_{max}}a^{-\frac{3i+1}{2}}f(a)da\Bigr) \\
\qquad\qquad\; = 3\cdot 10^{-3}\biggl[\beta(K)\left(\frac{a_0-1}{2a_0}+
\ln\frac{a_{max}}{a_0}\right)\biggr ]^{-1} \\
\quad\times M_l^{1/2}
\Biggl \{\frac{1}{a_0(a_0-1)}\left(\frac{2}{3}a_0^{3/2}-2a_0^{1/2}+
\frac{4}{3}\right) \\
\quad + 2a_0^{-1/2}- 1.35(M_l+M_s)^{-1/6}\\
\quad +\sum\limits_{i=1}^n(-1)^{i+1}
(3.24)^i(M_l+M_s)^{\frac{i}{2}}C_n^i\frac{2}{3i+1}\\
\quad\times \biggl[(2.2)^{-\frac{3i+1}{2}}
(M_l+M_s)^{-\frac{3i+1}{6}}
-a^{-\frac{3i+1}{2}}_{max}
\biggr]\Biggr \}. \hspace*{1.1cm}(22^{\prime})
\end{array}
$$

  We use the value of $0\fm 2$ for $K_0$ for all systems since it will be
shown in Section 5 that flares with such amplitudes are detectable
with the present-day observing facilities.

\begin{figure}[h]
\centering
\includegraphics[width=7.5cm]{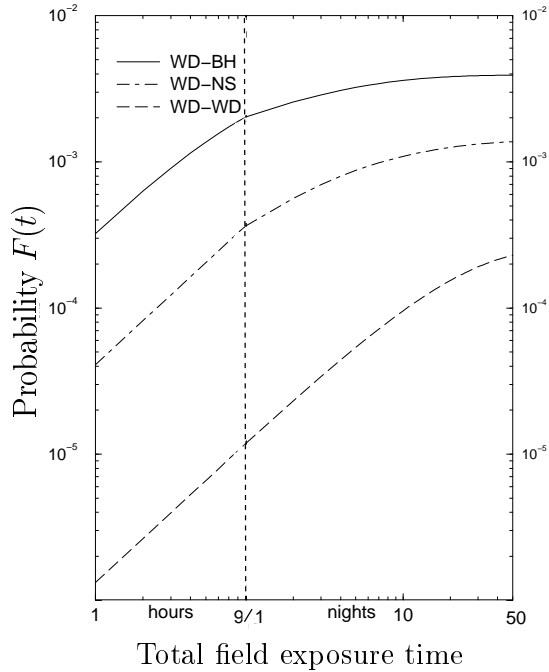}
\caption{Probability $F(t)$ of detecting a flare with $\Delta m\ge 0\fm 2$
as a function of the overall observation time of a specified sky region
measured either in hours or in nights.}
\end{figure}

The values of $F(K_0,t)$ for $K_0=1.2(\Delta m =0\fm 2)$ and $t=9^h$
are presented in Table 4 for a range of $a_0$ and $a_{max}$ values.
These rough model estimates may change by a factor of 1.5--2 depending on
the real pattern of the distribution of pairs of compact objects along the
semiaxes.

\begin{table*}
\centering
\caption[]{Probabilities of detection of flares with $\Delta m = 0\fm 2$ in 9 hours}
\vspace*{0.3cm}
\begin{tabular}{clcccccc}
	    \hline
	    \noalign{\smallskip}
&Parameters&\multicolumn{2}{c}{WD + WD}&\multicolumn{2}{c}{WD + NS}&\multicolumn{2}{c}{WD + BH}\\
&$\quad (R_{\odot})$    &            &               &  \\
\hline
\noalign{\smallskip}
&$a_{0}\;\;\;\setminus \;\;\; a_{max}$&50 & 30& 50 & 30& 50 & 30\\
\noalign{\smallskip}
\hline
\noalign{\smallskip}
&10 &$1.4\cdot 10^{-5}$&$1.6\cdot 10^{-5}$&$1.7\cdot 10^{-4}$&$2.2\cdot 10^{-4}$&$9.8\cdot 10^{-4}$&$1.3\cdot 10^{-3}$\\
&8 &$1.3\cdot 10^{-5}$&$1.4\cdot 10^{-5}$&$2.2\cdot 10^{-4}$&$2.7\cdot 10^{-4}$&$1.2\cdot 10^{-3}$&$1.6\cdot 10^{-3}$\\
&7 &$1.2\cdot 10^{-5}$&$1.3\cdot 10^{-5}$&$2.4\cdot 10^{-4}$&$3.1\cdot 10^{-4}$&$1.4\cdot 10^{-3}$&$1.8\cdot 10^{-3}$\\
&5.88&$1.1\cdot 10^{-5}$&$1.2\cdot 10^{-5}$&$3.0\cdot 10^{-4}$&$3.7\cdot 10^{-4}$&$1.6\cdot 10^{-3}$&$2.0\cdot 10^{-3}$\\
&4 &$9.9\cdot 10^{-6}$&$1.0\cdot 10^{-5}$&$4.2\cdot 10^{-4}$&$5.0\cdot 10^{-4}$&$2.2\cdot 10^{-3}$&$2.7\cdot 10^{-3}$\\
&3 &$9.1\cdot 10^{-6}$&$9.2\cdot 10^{-6}$&$4.9\cdot 10^{-4}$&$5.9\cdot 10^{-4}$&$2.6\cdot 10^{-3}$&$3.1\cdot 10^{-3}$\\
\noalign{\smallskip}
\hline
\noalign{\smallskip}
&Average&\multicolumn{2}{c}{$ 10^{-5}$}&\multicolumn{2}{c}{$4\cdot 10^{-4}$}&\multicolumn{2}{c}{$2\cdot 10^{-3}$}\\
&estimate
	   &    &  &  &&&\\
\hline
\end{tabular}
\end{table*}

Fig.7 represents the behaviour of $F(t)\equiv F(K_0,t)$ (for $\Delta m =
0\fm 2)$ as a function of $t$ and $n$ for the values of $a_0$ and $a_{max}$
which give approximately the average values among those presented in Table 4
for different system types.

We have also estimated the average durations and periods of reccurence
of the events considered and their dispersions using (20) -- ($22^{\prime}$).
To do this correctly one has to convolve $\tau$ and $T$ not with the $f(a)$
but with the $P(K_0,t,a)f(a)/F(K_0,t)$ according to the share each parameter
region brings to the probability of detecting a flare. These values,
formally, depend on the values of $t$ or $n$. However, it is clear that
their variations cannot influence much the real average characteristic.
Actually in these calculations we use the expression for $P(K_0,t,a)$ and
$F(K_0,t)$ in the region where they are linear in $t$,
which as we will see in Section 5 corresponds to the optimal strategy
for searching for the flares.

The results are presented in Table~5 for the value of amplitude
$\Delta m\geq 0\fm 2$.
The parameters averaged over the range of the used $a_0$ (from $3R_{\odot}$
to $10R_{\odot}$) are in the first and third columns. Variations of flare
duration with varying $a_0$ do not exceed a factor of 2. In columns 2 and
4 the limits of the intervals which contain parameters of 90~\% of flares
are given. The minimum values of the parameters correspond to the closest
systems of all types which are capable of giving rise to a flare with the
amplitude $\Delta m\ge 0\fm 2$. In Table~5 we give parameters of
events that one can hope to discover in observations.

\section{The number of pairs being investigated in the Galaxy}

As we have already noted, the present-day numbers of different pairs of
compact objects are determined in the framework of different versions of
population synthesis (Lipunov et al. 1996; Wellstein \& Langer 1999;
Fryer et al. 1999; Nelemans et al. 2001). However, the main objective
of this research is the study of systems which are of interest within the
scope of gravitational-wave astronomy and of the problem of origin of
gamma-ray bursts. These are pairs of neutron stars, neutron stars with black
holes and essentially rarely white dwarfs with black holes. By now
extensive work has been done on the modeling of the origin and evolution of
pairs of white dwarfs. It goes without saying that these binary systems
are best understood; comparison with observations can be made for them
(Hernanz et al. 1997; Nelemans et al. 2001). There are at least two
relatively reliable estimates of the total number of binary white dwarfs
in the Galaxy, $5\cdot 10^{8}$ (Lipunov et al. 1996) and $2.5\cdot 10^8$
(Nelemans et al. 2001). In principle, these data are sufficient
to determine the number of gravitational-lens flares that can be detected
in this sample. Nevertheless, we have estimated the number of binary
white dwarfs and also white dwarfs in pairs with neutron stars and black
holes in the Galaxy in the framework of assumptions of  the progenitors of
binary systems consisting of compact objects. We have followed Bethe and
Brown (1998, 1999) who have determined the number of massive binaries
containing neutron stars and black holes.

\begin{table}[h]
\centering
\caption[]{Parameters of flares of amplitude $\Delta m\ge 0\fm 2$
}
\vspace*{0.3cm}
\begin{tabular}{@{}l@{}ccc@{}c@{}}
	    \hline
	    \noalign{\smallskip}
&\multicolumn{2}{c}{Duration (s)}&\multicolumn{2}{c}{Period of recurrence (h)}\\
&average &interval of  & average &interval of \\
&        &  90\%       &         & 90\%\\
\hline
\noalign{\smallskip}
WD+BH &50       & 10-135& 15 & 1-48\\
WD+NS &60       & 20-120& 42 & 8-120\\
WD+WD &135 	& 90-180& 400& 120-500\\
\noalign{\smallskip}
\hline
\end{tabular}
\end{table}

We will base our discussion on the following assumptions (as previously,
solar units are used).

1. The mass distribution of progenitor stars (initial mass function)
obeys Salpeter's (1955) law in the interval $0.1 < M < 120$, i.e.
\begin{equation}
dW(M) = w(M)dM = 1.35(\frac{M}{0.1})^{-2.35}dM\; ,
\end{equation}
where $dW(M)$ is the probability that the mass of a star will be in the
interval $M\div M + dM$, $w(M)$ is the probability density.

At the observed star formation rate $\frac{dN}{dM\cdot dt} =
0.9(\frac{M}{0.1})^{-2.35} {\rm year}^{-1}$ and the Galaxy age of $10^{10}$
years, the number of stars is $N\sim 1.5\cdot 10^{11}$.

2. About 100~\% of stars are members of binary systems. This evaluation
is used in the population synthesis as well as 50~\% (see for instance,
Nelemans et al. 2001).

3. The distribution density in the mass ratio, $q\equiv M_s/M_l\le 1$,
is constant, i.e. $\frac{dN}{dq}\propto {\rm const}$ ( e.g. Portegies Zwart
\& Yungelson 1998).

4. The initial masses of progenitor stars must lie within the following
intervals to give: \\
white dwarf -- $1 < M < 10$,
neutron star -- $10 < M < 25$,
black hole -- $25 < M$.\\
In the last case the uncertainty is rather great; the values
$40 M_{\odot}$ (Van den Heuvel \& Habets 1984), and $60 M_{\odot}$
(Woosly et al. 1995) are also used. As for us, we follow Portegies Zwart
et al. (1997).

5. The neutron star formed in a supernova explosion in a binary system
receives a kick. As Cordes \& Chernoff (1997) have shown,
its distribution is well approximated by the sum of two Gaussians with
standard 175 km/s and  700 km/s, the share of the former being 80~\% while
that of the latter being 20~\%. With these parameters 43~\% of binary systems
survive after the explosion (Bethe \& Brown 1998). As a black hole is formed,
one can believe that the kick will be lower, at least, inversely proportional
to its mass (Fryer et al. 1999). For this case, using the relationships
between the probability of decay and the additional velocity derived by Bethe
\& Brown (1998), we have found that 87~\% of binaries survive after the
formation of a black hole.

Now we turn to estimation of the number of pairs of different types.

For the probability $d\phi$ to obtain a binary system with the mass of the
more massive companion (in our case the object is a lens) in the interval
$M_l\div M_l + dM_l$ and the mass ratio in the interval $q\div q + dq$,
considering (23) and the homogeneity of the distribution in $q$ ($M_l$ is
measured in $0.1 M_{\odot}$), we have
\begin{equation}
d\phi = dqdW = 1.35(M_l)^{-2.35}dqdM_{l}.
\end{equation}
The number of pairs of type $i$ is $N_i = \phi_i\gamma_i N_{bin}$, $i = d$
for a pair of white dwarfs, $i = n$ and $i = b$ for binaries containing a
white dwarf with a neutron star and a black hole, respectively; $\gamma_i$
is the pair survival probability after the supernova explosion.

Clearly
\begin{equation}
\phi_i = \int\limits_{Q_i}\int\limits_{M_i}d\phi = 1.35\int\limits_{Q_i}\int
\limits_{M_i}(M_l)^{-2.35}dqdM_{l},
\end{equation}
where $Q_i$ is the interval where the values of $q$ lie, and $M$ is the
interval of $M_l$ values. Determine $Q_i$ for each type of pairs, taking
into account that $10 < M_s = qM_l < 100$ (assumption 5).
Then $\frac{10}{M_l} <
q < 1$ for a pair WD--WD, $\frac{10}{M_l} < q < \frac{100}{M_l}$
for a pair WD--NS and WD--BH. Thus, proceeding from (25), we obtain for a
pair of white dwarfs
\begin{equation}
\begin{array}{ll}
\phi_d& = 1.35\int\limits_{10}^{100}(M_l)^{-2.35}
dM_{l}\int\limits_{\frac{10}{M_l}}^{1}dq \\
& = 1.35\int\limits_{10}^{100}(M_l)^{-2.35}
\left(1 - \frac{10}{M_l}\right )dM_{l}.
\end{array}
\end{equation}
Hence
$$
\begin{array}{ll}
\phi_d& = 1.35\int\limits_{10}^{100}M_l^{-2.35}dM_l - 13.35
\int\limits_{10}^{100}M_l^{-3.35}dM_l  \\
& = 0.043 - 0.025 = 0.018,
\end{array}
$$
and allowing for the fact  that the number of pairs in the Galaxy is
$N_{bin} = 0.5N$, obtain for the number of binary white dwarfs
$N_d = 1.35\cdot 10^9$ pairs.

For white dwarfs in pair with neutron stars
$$
\begin{array}{ll}
\phi_n& = 1.35\int\limits_{100}^{250}M_l^{-2.35}dM_l
\int\limits_{\frac{10}{M_l}}^{\frac{100}{M_l}}dq \\
& = 1.35\int\limits_{100}^{250}M_l^{-2.35}\frac{90}{M_l}
dM_l = 90\cdot 1.35\int\limits_{100}^{250}M_l^{-3.35}dM_l,
\end{array}
$$
and finally
$$
\phi_n = \frac{90\cdot 1.35} {2.35}\left [M_l^{-2.35}\Bigr |_{100}^{250}
\right ]= 51.7\cdot 1.76\cdot 10^{-5} = 9.1\cdot 10^{-4}.
$$
Because 57~\% of binary systems decay after the supernova explosion which
results in formation of a neutron star, $N_n\sim 9.1\cdot 10^{-4}\cdot
0.43\cdot 0.5N\sim 2.9\cdot 10^7$ pairs.

The same is for pairs containing black holes:
$$
\phi_b = 51.7\left [M_l^{-2.35}\Bigr |_{250}^{1200}\right ] = 51.7\cdot 2.26
\cdot 10^{-6} = 1.2\cdot 10^{-4}.
$$
Since the probability of their survival is 87\%, then $N_b\sim 1.2\cdot
10^{-4}\cdot 0.87 \cdot 0.5\cdot N\sim 7.8\cdot 10^6$ pairs.

It should be emphasized that the above-given estimates are defined in many
respects by the choice of the power index $k$ in the initial function of
stellar masses. Clearly they decrease with increasing $k$. It is likely
that the version with $k = 2.35$ of Salpeter (1955) is most consistent with
current observational data (see, for instance, Grishchuk et al. 2001;
Raguzova 2001), but, nevertheless, $k = 2.5$ and also $k = 2.7$ are used
in the population synthesis as well (Bethe \& Brown 1998, 1999; Nelemans
et al. 2001).

To make the picture complete, we present in Table~6 the estimates of the
numbers of pairs in the Galaxy for the mentioned $k$.

\begin{table}[h]
\centering
\caption[]{Estimates of the number of pairs of compact objects for
different mass functions}
\vspace*{0.3cm}
\begin{tabular}{cccc}
	    \hline
	    \noalign{\smallskip}
 $k$     &WD--WD     &WD--NS&WD--BH\\
	 &            &       &  \\
\hline
\noalign{\smallskip}
2.35 &$1.3\cdot 10^9$&$2.7\cdot 10^7$ &$8\cdot 10^6$\\
2.5  &$8.7\cdot 10^8$&$1.6\cdot 10^7$ &$4\cdot 10^6$\\
2.7  &$5.5\cdot 10^8$&$6\cdot 10^6$   &$1.2\cdot 10^6$ \\
\noalign{\smallskip}
\hline
\end{tabular}
\end{table}
On the other hand, in order to determine the accuracy of our estimates,
one can compare them (for binary white dwarfs) with the results of population
synthesis. In particular, the number of pairs of this type is $4.5\cdot 10^8$
(Lipunov et al. 1996) and $2.5\cdot 10^8$ (Nelemans et al. 2001). Note
that the last number was obtained for $k = 2.7$ and it is 2 times as low as
our rough value from Table~6. The 2-4--time differences are reasonable,
taking into account the uncertainty in the results of modeling of evolution
and the qualitative character of our estimates.

\section{Strategy of search and the number of detectable objects.
Conclusions}

To determine the number of objects that may be detected using a certain
telescope one has to know not only the overall number but also their
distribution in the Galaxy, their luminosity function and the distribution
of absorbing matter as well.

	We used the star distribution in the Galaxy of the following form:
\begin{equation}
	\rho=\rho_0e^{-\frac{R}{H}}e^{-\frac{|z|}{h}} ,
\end{equation}
    where $R$ and $z$ are the point cylindrical coordinates in the Galaxy,
$H$ and $h$ are radial and vertical distance scales, respectively, for
which we
used the following values: 
$H=8$~kpc and $h=250$~pc (see Dehnen \& Binney 1998; Nelemans et al. 2001).
Adopting the Sun's coordinates $R_{\odot}=8.5$~kpc and $z_{\odot}=30$~pc
we have:
$$
	\rho_0=\rho_{loc}e^{\frac{R_{\odot}}{H}}e^{\frac{z_{\odot}}{h}}.
$$

The overall number of objects in the Galaxy is expressed by the integral
$$
N=\int\limits_{-\infty}^{\infty}dz\int\limits_0^{\infty} 2\pi R \rho_0e^
{-\frac{R}{H}}e^{-\frac{|z|}{h}} dR,
$$
	which yields
\begin{equation}
\rho_{loc}=\frac{Ne^{-\frac{R_{\odot}}{H}}e^{\frac{|z_{\odot}|}{h}}}
{4\pi H^2h} =\frac{N}{6.56\cdot 10^{11}} {\rm pc^{-3}}
\end{equation}
	and
\begin{equation}
\rho(R,z)=\frac{N}{6.56\cdot 10^{11}}e^{-\frac{R-R_{\odot}}{H}}e^
{-\frac{|z|-z_{\odot}}{h}} {\rm pc^{-3}}.
\end{equation}
We used distributions for WD-NS and WD-BH in the same form with the
only difference in normalization coming from their fewer overall number in 
the Milky Way.
	The local densities corresponding to the overall numbers given 
in Table 6 are presented in Table 7.

Note that the part of WD-WD pairs among the complete number of
WDs is (7-25)~\% (Iben et al. 1997; Fryer et al. 1999) and the WD
density in the Solar neighbourhood is $(4-20)\cdot 10^{-3}$ pc$^{-3}$
(see Nelemans et al. 2001 and references therein). This gives us a range of
local density of $3\cdot 10^{-4} - 5\cdot 10^{-3}$~pc$^{-3}$. It is very
close to our estimate shown in Table 7 and the most optimistic values of
density may be $2-2.5$ times more than those.

 Unfortunately, there is no data to compare with for binaries other
than WD-WD. We suggest only that our estimation range may be $2-3$ times
narrower. Therefore, the real number of objects may be a few times higher.
\begin{table}[h]
\centering
\caption[]{The local densities of objects derived from Table 6 data
and distirbution (27) (in pc$^{-3}$).}
\begin{tabular}{cccc}
	    \hline
	    \noalign{\smallskip}
 $k$     &WD--WD     &WD--NS&WD--BH\\
	 &            &       &  \\
\hline
\noalign{\smallskip}
2.35 &$2.0\cdot 10^{-3}$&$4.1\cdot 10^{-5}$ &$1.2\cdot 10^{-5}$\\
2.5  &$1.3\cdot 10^{-3}$&$2.4\cdot 10^{-5}$ &$6.1\cdot 10^{-6}$\\
2.7  &$8.4\cdot 10^{-4}$&$9.1\cdot 10^{-6}$ &$1.8\cdot 10^{-6}$\\
\noalign{\smallskip}
\hline
\end{tabular}
\end{table}

 The insterstellar absorption is proportional to the interstellar medium
density which is distributed in the Galaxy by the same~ (27)
law with $h=80$~pc (Dehnen \& Binney 1998). The normalization was done
according to the local average value of absorption $A_{g^{\prime}}=1\fm 16
{\rm kpc^{-1}}$ in the $g^{\prime}$ band of the SDSS equipment
(Schlegel et al. 1998).

 The WD luminosity function is well known and is explained fairly well
in the frame of their cooling theory. We used the luminosity function of
Oswalt et al. (1996). It is readily seen from this that more than half
of the WDs are brighter than $13\fm 5$ and only less than 20 percent are
fainter than $15^{\rm m}$, where the function reaches its maximum. Thus, the
magnitudes of the majority of WDs do not suffer much from the bolometrical
corrections which is a few tenth for the stars of these spectral
classes (see, e.g. Allen 1973).

In other binary systems, the situation is more complicated. It is only clear
that the optical emission from both a NS and a BH can be associated with
interstellar gas accretion. In the former case it is likely to be negligibly
small because of the high orbital velocity of the NS ($> 100$ km/s)
(Shapiro \& Teukolsky 1983). At the same time the orbital velocity of a BH
in a pair with a WD is about 50 km/s, and the luminosity of accretion plasma
can reach $10^{30}$ erg/s by moderate-optimistic estimates (Shvartsman 1971;
Beskin \& Karpov 2002) and will be by an order of magnitude less by a
pessimistic one (Ipser \& Price 1982). Apparently, its contribution to the
total luminosity of the system is insufficient to correct the depth of the
space of detection. However, this level of optical emission makes possible
searching for its fast variations with a microsecond time resolution for
investigation of accretion processes and strong gravitational fields near
the horizon of events (Shvartsman 1971; Beskin et al. 1997;
Beskin \& Karpov 2002).

Now we have everything we need to calculate the on-sky density of objects
in a given direction ($l,b$):
\begin{equation}
\begin{array}{ll}
	\sigma(l,b)&=\frac{d^2N}{\cos{b} dl db}\\
	&=\int r^2 \rho(R,z) I(M)dr\; ,
\end{array}
\end{equation}
where $I(M)$ is a cumulative luminosity function from Oswald et al.
(1996), $M=m_{lim}-5\lg(r/10)-A(r,l,b$), $m_{lim}$ is an equipment
limiting apparent stellar magnitude, $A(r, l, b)$ is a total extinction
in the $(l,b)$ direction up to the distance $r$, and 
$$
\begin{array}{l}
	z=z_{\odot}+r\sin{b},\\
        R=\sqrt{R^2_{\odot}+r^2\cos^2{l}-2R_{\odot}r\cos{b}\cos{l}}
\end{array}
$$
  	are the point cylindrical coordinates expressed in terms of its
galactic coordinates $(l,b)$ and distance $r$.

  Unfortunately, this integration cannot be done analytically and so we
will turn at once to a particular version of the equipment being discussed.
In our view, an ideal tool for carrying out the proposed programme is
currently the  2.5-metre telescope at APO (New Mexico).
It has been used to accomplish one of the most promising projects --
the Sloan Digital Sky Survey (SDSS). The telescope has a field of
$3^{\circ}$, in which 30 CCD chips of $2048\times 2048$ pixels
are mounted (York et al. 2000). Sky scanning at a speed of its movement along
a strip $2\fdg 5$ wide is performed within the framework of the
project. During an exposure of 55 s, the limiting stellar magnitude in
the $g^{\prime}$ filter (it is close to the $B$ filter of the Johnson
system) is $23\fm 2$ at a signal-to-noise ratio of $\sim 5$
(Oke \& Gunn 1983; York et al. 2000). We have constructed a relationship
between amplitude of flares recorded at a $5\sigma$ level and stellar
magnitude of a quiet object in the $g^{\prime}$ band (Fig.~8). In so doing,
we used the parameters corresponding to the  SDSS project: the seeing --
0.8 arcsec, the quantum efficiency of the CCD matrix -- 80~\%.

By comparing the data of Fig.~8 with the mean parameters of flares (Table~5)
and their light curves for different binary systems (Fig.~5), it is clear
that with a particular exposure of time $t_e\sim 10-30$ s,
the flares with $\Delta m\ge 0\fm 2$ of any pairs that we have examined can
be recorded at a significance level of $5\cdot 10^{-3} - 10^{-5}$ (duration of
any flare is a few $t_e$). In fact, the level of the flare detection will
be better due to registration in the SDSS of the field in several colour bands
($u^{\prime}, g^{\prime}, r^{\prime}, z^{\prime}$) and
the limiting $r^{\prime}$ stellar magnitude is also $23\fm$ Moreover, the
use and comparison of the data in the $g^{\prime} $ and $r^{\prime}$
filters give the possibility of easy separation of cosmic ray traces.

\begin{figure}[h]
\centering
\centering\includegraphics[width=8.5cm,
]{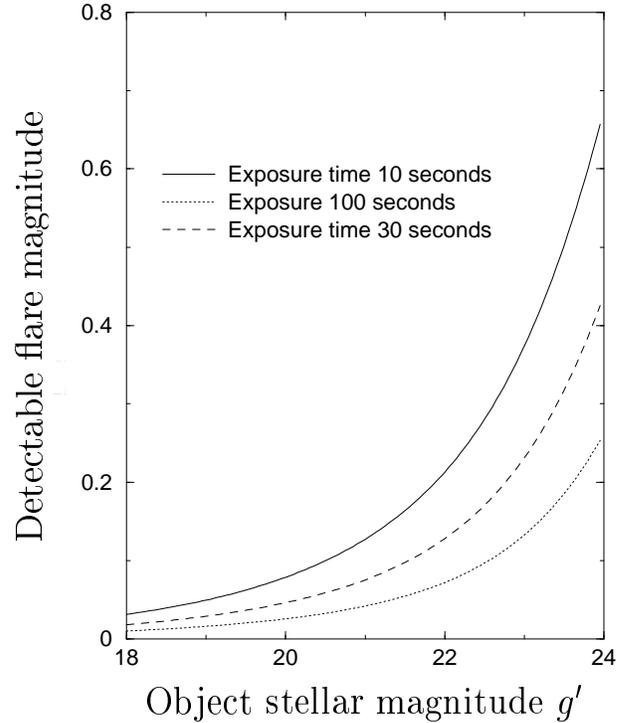}
\caption{Magnitudes of detectable flares in the SDSS versus object
brightness}
\end{figure}

On the basis of (30) we have built a map of the expected on-sky densities of
WD-WD pairs up to $m_{lim}=23^{\rm m}$ which is presented in Fig. 9. The
value of the local density corresponding to the Salpeter mass-spectrum index
of $2.35$ is used.
One can see from the map that the maximal density regions have galactic 
latitudes of $5-10$ degrees and a question might arise if it is possible to
observe in such star-rich regions of the sky. However, even in these latitudes
the average distance between stars of up to $23^{\rm m}$ is about 3-4
arcseconds (see e.g., Zombeck, 1982, p.34) and thus, they are resolved with
the SDSS seeing and this crowding therefore  cannot have much negative
impact on observations with this equipment.

\begin{figure}[h]
\centering
\centering\includegraphics[width=8.5cm]{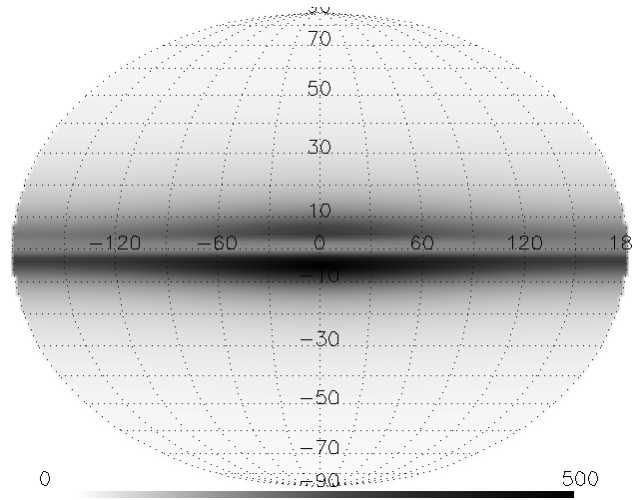}
\caption{Simulated on-sky distribution of WD-WD pairs, $m_{lim}=23^{\rm m}$.
Maximum density value is 490 pairs per square degree.}
\end{figure}

To derive the expected number of events of the discussed type  $N_{exp}$ one
multiplies the probability of detecting it from a randomly chosen pair
$F(t)$ for $\Delta m=0\fm 2$ (see (15), (20)-(22$^{\prime}$) and Fig.7)
by the number $D(\Omega)$ of such pairs observed during a certain
observational programme that covered the celestial sphere part $\Omega$:
\begin{equation}
	N_{exp}=F(t)\times D(\Omega) ,
\end{equation}
where
\begin{equation}
    D(\Omega)=\int\limits_{\Omega} \sigma(l,b) \cos{b} dl db\; .
\end{equation}

 It is clear, however, that when applying to reality one deals with
a telescope of limited field of view and the observational time provided for 
a programme is limited as well, so increasing the total exposure of a field 
to increase $F(t)$ leads to decreasing the number of such fields in the
programme. It is also evident that one has to monitor at first the best, i.e.
maximal density $\sigma(l,b)$ fields first. Fig. 10 represents the number of
pairs of each type $D(\Omega)$, numerically calculated from (32),
in $\Omega$ fields with a size of $2\fdg 5\times 2\fdg 5$ (field of view of
SDSS) after sorting them in decending order of pair density.

\begin{figure}[h]
\centering
\centering\includegraphics[width=8.5cm,
]{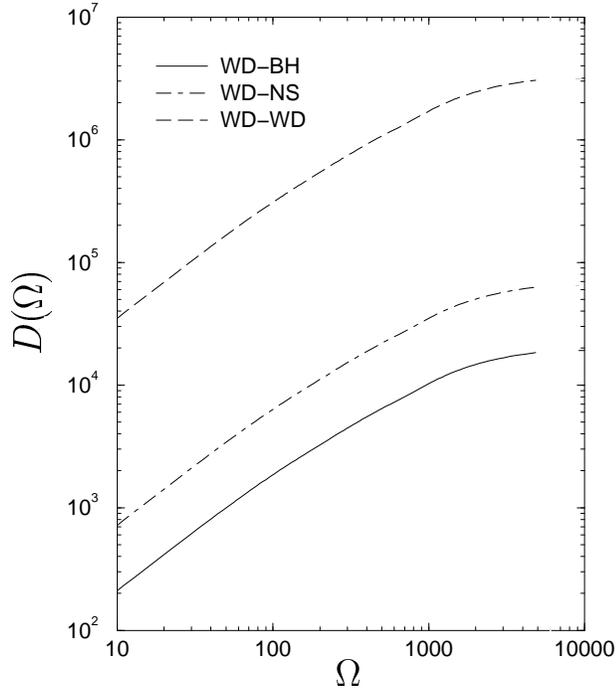}
\caption{Expected number of objects $D(\Omega)$ in $\Omega$
the best (with the highest density of pairs)
$2\fdg 5\times 2\fdg 5$ fields.}
\end{figure}

 Now let the programme last for 5 years. Considering that the matter
in question is detection of rather faint objects, observations have to be
made on dark nights only at moon phases within $\pm 5^d$ from the new moon,
which gives approximately $36 \%$ of night time. Assuming the average
observational night of $9^h$ the 5-year resource of time then will make
$n_{obs}=657$ nights or $T_{obs}=5913$ hours. Thus, as every field is
observed for $t$ hours or $n$ nights (and $\Omega = T_{obs}/t$, $n_{obs}/n$)
the expected number of detected pairs is
\begin{equation}
	N_{exp}=D\left(\frac{T_{obs}}{t}\right) F(t),
\end{equation}
	or
$$
	N_{exp}=D\left(\frac{n_{obs}}{n}\right) F(n).
\hspace*{4.6cm}(33^{\prime})
$$

These quantities derived from Figs.~7 and 10 are presented in
Fig.~11. One can see from this figure that the optimal total exposure
per field for WD-WD pairs is $6-7$ nights, and it is about 1 night when
searching for WD-NS or WD-BH systems.

\begin{figure}[t]
\centering
\centering\includegraphics[width=8.5cm,
]{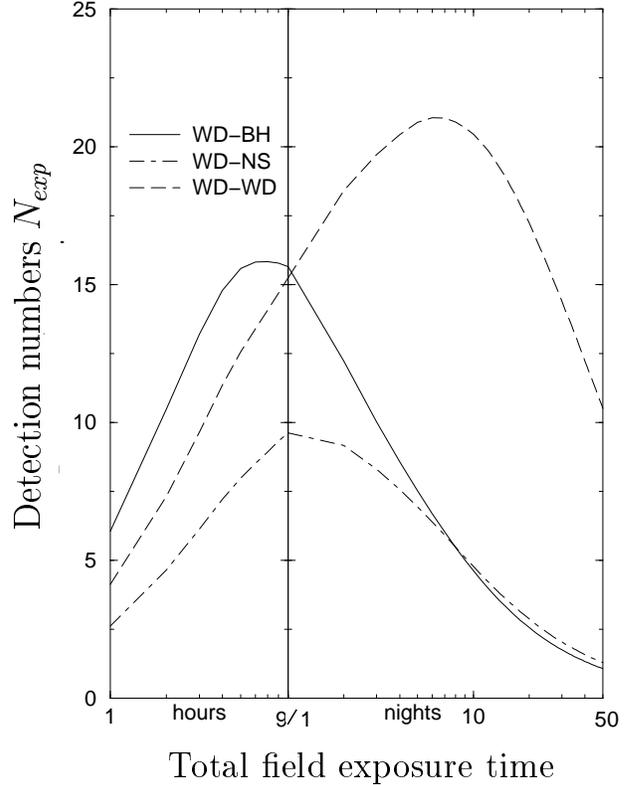}
\caption{Expected detection numbers in the 5-year programme as a function of
total exposure per field.}
\end{figure}

 The expected numbers of detections with optimal strategies along with
the optimal total exposures per field are given at Table 8.
\begin{table}[h]
\centering
\caption[]{The number of pairs of compact objects detectable over 5 years
and the optimal total exposures per field}
\begin{tabular}{lccc}
	    \hline
	    \noalign{\smallskip}
 $k$     &WD--WD     &WD--NS&WD--BH\\
	 &            &       &  \\
\hline
\noalign{\smallskip}
2.35 &22&9&16\\
2.5  &15&5&8\\
2.7  &9&2 &3 \\
\hline
Optimal & & & \\
exposure time&6-7 nights &1-2 nights &6-9 hours \\
per $2\fdg 5\times 2\fdg 5$ field&&&\\
\noalign{\smallskip}
\hline
\end{tabular}
\end{table}

A new project entitled "The Dark Matter Telescope"
({\tt http://www.dmtelescope.org/index.htm}) being developed involved
8.4 (effectively 6.9) metre telescope with a field of view slightly wider than
that of the SDSS. The use of it in a manner similar to that described above
increases the number of detectable objects by a factor of 2.2. Other wide
field telescopes of 4 metres in diameter being built now, LAMOST
({\tt http://www.greenwich-observatory.co.uk/lamost.\\html}) and VISTA
({\tt http://www.vista.ac.uk}), could detect 1.5 times more binaries
than the SDSS.

Thus, for searching for the flares caused by gravitation lensing in binary
systems with compact companions, the facilities of the SDSS can
be used, having reduced the time of an individual exposure to
10-20 s. It is obvious that after recording a flare in a certain
object, another telescope of similar class must be involved in observations.
This object should be monitored to prove the effect and for detailed
investigation of the detected system.

Thus, within the frame of the proposed programme, it will be possible to
find several dozen binary systems comprised of compact objects. It will be
recalled that in numerous surveys of microlensing, the number of reliably
detected events amounts to about 350 (Alcock 1999),
only in two cases the matter in question being of massive lenses
(black holes?) (Bennett et al. 2001). When applying the technique suggested,
there is a possibility of detecting with absolute assurance more than
10 black holes with open event horizons. An investigation of these
objects with high time resolution within the MANIA experiment
(Beskin et al. 1997) may permit at last the detection of observational
evidence of extreme gravitational fields.

\begin{acknowledgements}
This investigation was supported by the Russian Ministry of Science,
Russian Foundation of Basic Researches (grant 01-02-17857),
Federal Programs "Astronomy" and "Integration" and  Science-Education
Center "Cosmion".
The authors are very grateful to the anonymous referee for very useful
observations and E. Agol for important notes.
AVT would like to thank D.A.Smirnov and S.V. Karpov for valuable discussions
and his especially grateful to I.V. Chilingaryan for his help with various
questions. We thank T.I. Tupolova for manuscript preparation.

\end{acknowledgements}

\end{document}